\documentclass[a4paper]{article}
\usepackage{epsfig,amsmath,amsfonts,amssymb,setspace,multirow,textcomp}
\usepackage[T1]{fontenc}
\makeatletter
\renewcommand{\@oddhead}{\textit{Advances in Astronomy and Space Physics} \hfil}
\renewcommand{\@evenfoot}{\hfil \thepage \hfil}
\renewcommand{\@oddfoot}{\hfil \thepage \hfil}
\makeatother

\renewenvironment{thebibliography}[1]{\begin{oldthebibliography}{#1}\setlength{\parskip}{0ex}\setlength{\itemsep}{0ex}}{\end{oldthebibliography}}

\begin{document}
\fontsize{11}{11}\selectfont 
\title{High-speed multicolor photometry with CMOS cameras}
\author{\textsl{S.\,M.~Pokhvala$^{1}$, B.\,E.~Zhilyaev$^{1}$, V.\,M.~Reshetnyk$^{2}$}}
\date{\vspace*{-6ex}}
\maketitle
\begin{center} {\small $^{1}$Main Astronomical Observatory, NAS of Ukraine, Zabalotnoho 27, 03680, Kyiv, Ukraine\\
$^{2}$Taras Shevchenko National University of Kyiv, Glushkova ave., 4, 03127, Kyiv, Ukraine\\
{\tt nightspirit@ukr.net}}
\end{center}

\begin{abstract}
We present the results of testing the commercial digital camera
Nikon D90 with a CMOS sensor for high-speed photometry with a small
telescope Celestron 11" on Peak Terskol. CMOS sensor allows to
perform photometry in 3 filters simultaneously that gives a great
advantage compared with monochrome CCD detectors. The Bayer BGR
color system of CMOS sensors is close to the Johnson BVR system. The
results of testing show that we can measure the stars up to V
$\simeq$ 14 with the precision of 0.01 mag.  Stars up to magnitude V
$\sim$ 10 can shoot at 24 frames per second in the video mode.


\end{abstract}

\section*{\sc introduction}

\indent \indent Synchronous multicolor observations are important in
many cases, for instance, for the study of transit of extrasolar
planets, afterglow of gamma-ray burst, small-scale variations during
flares on dwarf stars and cataclysmic variables and many others.
Fast simultaneous photometry in several bands is especially
important to obtain information about the spectra of radiation at
short time intervals during transient events. Traditional methods
lose a signal due to the serial measurements in different filters.
Multicolor sensors based on metal-oxide semiconductors (CMOS) allow
to refuse the use of filters in general \cite{butler08}. In recent
years, CMOS sensors have made a serious competition to CCD. CMOS
sensors provide simultaneous imaging of the object in the Bayer
color system: the blue filter "B", ($\lambda$ > 400 nm), the red "R"
($\lambda$ <900 nm) and the intermediate filter "G". Transformation
of the Bayer color system BGR to the international Johnson BVR
system is a relatively simple problem. CMOS sensors allow
non-destructive reading of digital images. Currently, algorithms are
developed that allow achieve a high dynamic range, to avoid
blooming, to correct tracking errors of the telescope, the variation
of atmospheric transparency \cite{butler08}. The disadvantage of
CMOS sensors is relatively low quantum efficiency. However, it plays
a role only under observations of extremely faint stars. Commercial
CMOS sensors allow us to perform fast high-precision multicolor
photometry of relatively bright objects with small telescopes. Using
high values of ISO, 6400 and above, up to ISO 25600 provides a great
opportunity to capture objects up to V $\sim$ 10 magnitude with a
frequency up to 24 frames per second and higher with small
telescopes. This allows us to study high-frequency variability in
the range of 10 Hz and above.

\section*{\sc Limiting magnitudes}

\indent \indent Read noise and thermal noise are the main
controlling factors for detecting faint objects. Old CCDs have read
noise levels in the 15 to 20 or more electrons. Newer CCDs tend to
run in the 4 to 3 electron range.  As mentioned by Clark
\cite{clark12} the technology improvements of CMOS sensors have led
recently to that read noise dropped to about 2.5 electrons
(commercially available cameras Canon 5D Mark II, 50D, 7D). As
mentioned by Clark \cite{clark12} Nikon's technology currently
restricts the average read noise at zero level, losing some data.
Canon includes an offset, so processing by some raw converters can
preserve the low end noise, which can be important for averaging
multiple frames to detect very low intensity objects. The dark
current of CMOS image sensors is typically of the order of 60 - 100
electrons/sec at room temperature. In the latest models for
professional use (Fairchild Imaging CIS1021) it is $\sim$ 26
electrons/sec at 20 $ ^{\circ}$ C. It should be noted that Canon
develops CMOS image sensor independently. Table 1 shows that the
value of dark current of Canon EOS 20D at room temperature is less
than for Peltier cooled CCDs. It defines the boundary of the photon
mode when the signal from a star becomes comparable to the dark
current. The boundary of the photon mode can be found from the
expression for the illumination of the star image on the matrix
$E_{m}$:

\begin{equation}\label{form1}
E_{m} = C_{m}T_{a}T_{i} S/s, \,\,[lux]
\end{equation}
where $C_{m}$ is the illumination from a star of magnitude $m$,
$T_{a}$, and $T_{i}$ are the transmittance of the atmosphere and of
the instrument, $S$ and $s$ are the areas of telescope aperture and
stellar image.

Illumination of 1 $lux$  creates a flux of $\sim 5\cdot10^{11}$
$cm^{-2}\, s^{-1}$. Hence the expression for the flux through the
pixel is equal to
\begin{equation}\label{form2}
F_{m} = 5\cdot10^{11}E_{m}\,\varepsilon^{2}, \,\, [photon\, s^{-1}]
\end{equation}
where $\varepsilon$ is the pixel dimension.

As one can see from Fig. 1 dark current-limited boundary for the
Celestron 11" telescope equipped with Nikon D90 camera is around of
V $\simeq$ 18.


We analyzed some parameters of the professional astronomical sensors
and compared them with the professional commercial camera Canon EOS
20D (Table 1).

The processing of the observational data obtained at the observatory
Peak Terskol, confirmed results of our calculations. We can measure
the stars up to 14 mag with high accuracy, as shown in Fig. 2. We
can also observe stars up to 10 mag in video mode as shown in Fig.
3.

\begin{figure}[!h]
\centering
\begin{minipage}[t]{.45\linewidth}
\centering
\epsfig{file = 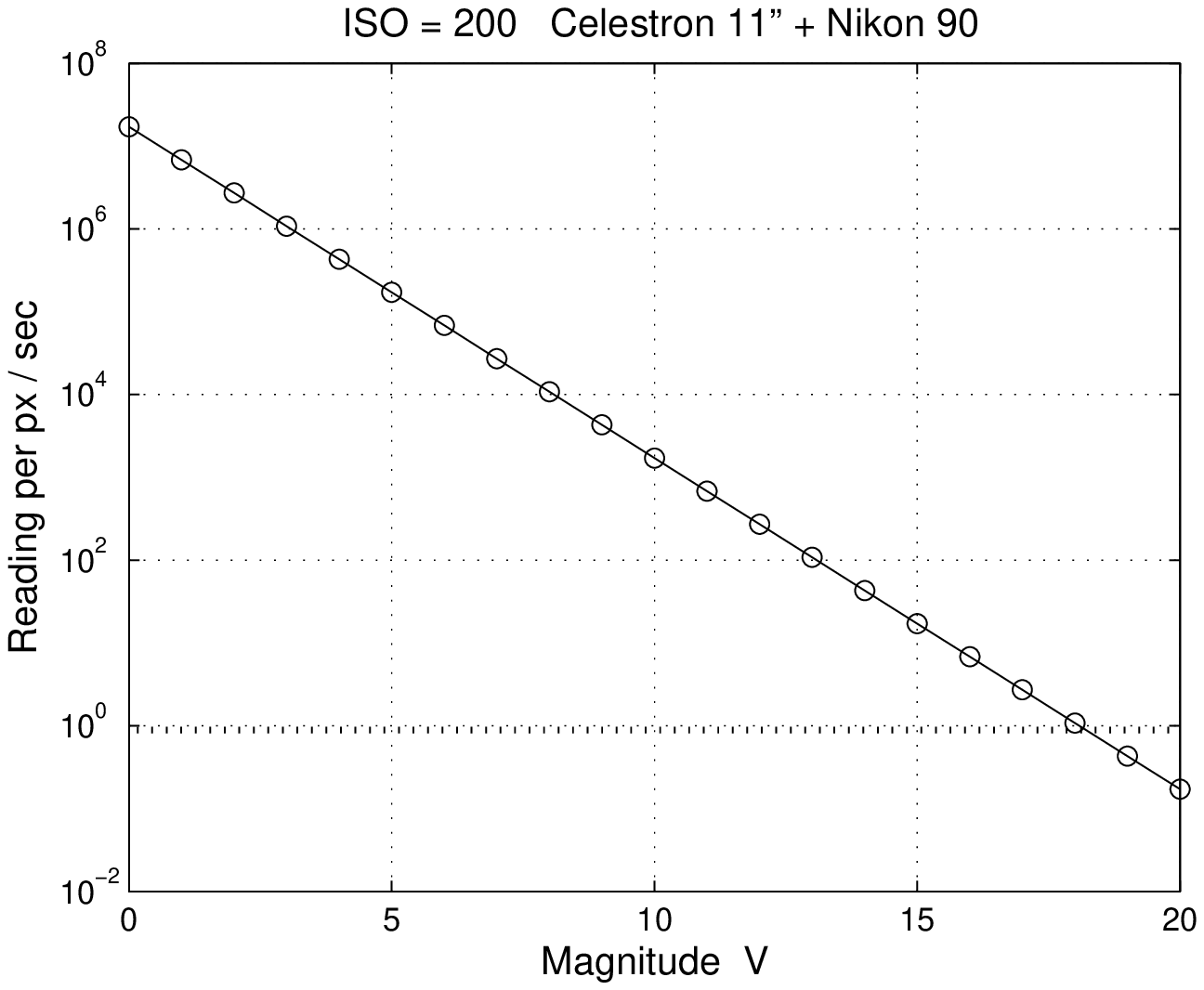,width = 1.0\linewidth} \caption{Dark
current-limited boundary is around V = 18.}\label{fig1}
\end{minipage}
\hfill
\begin{minipage}[t]{.45\linewidth}
\centering
\epsfig{file = 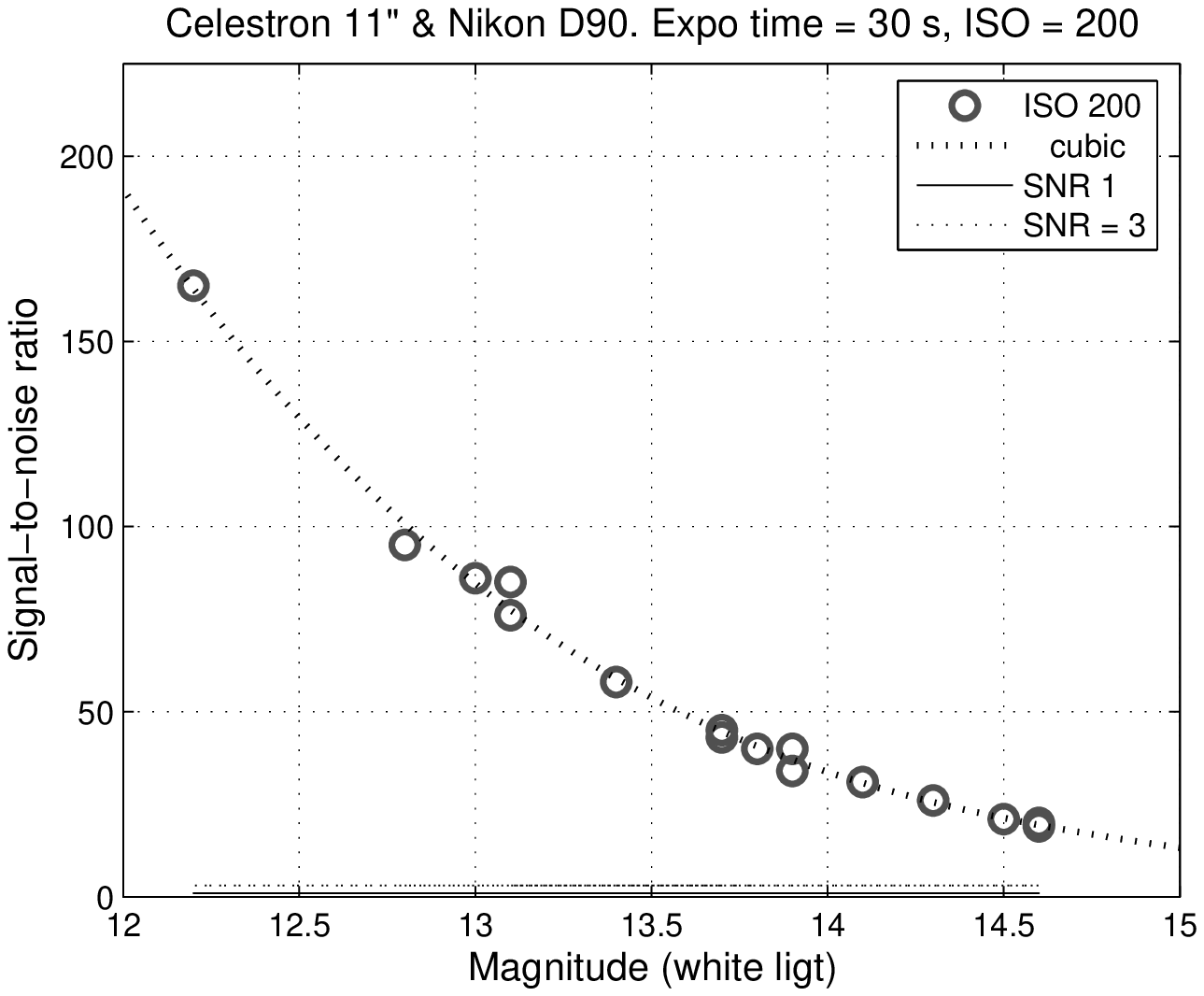,width = 1.0\linewidth} \caption{The
S/N ratio vs. B+G+R mag. The exposure time is 30 s.}\label{fig2}
\end{minipage}
\end{figure}

\begin{figure}[!h]
\centering
\begin{minipage}[t]{.45\linewidth}
\centering
\epsfig{file = 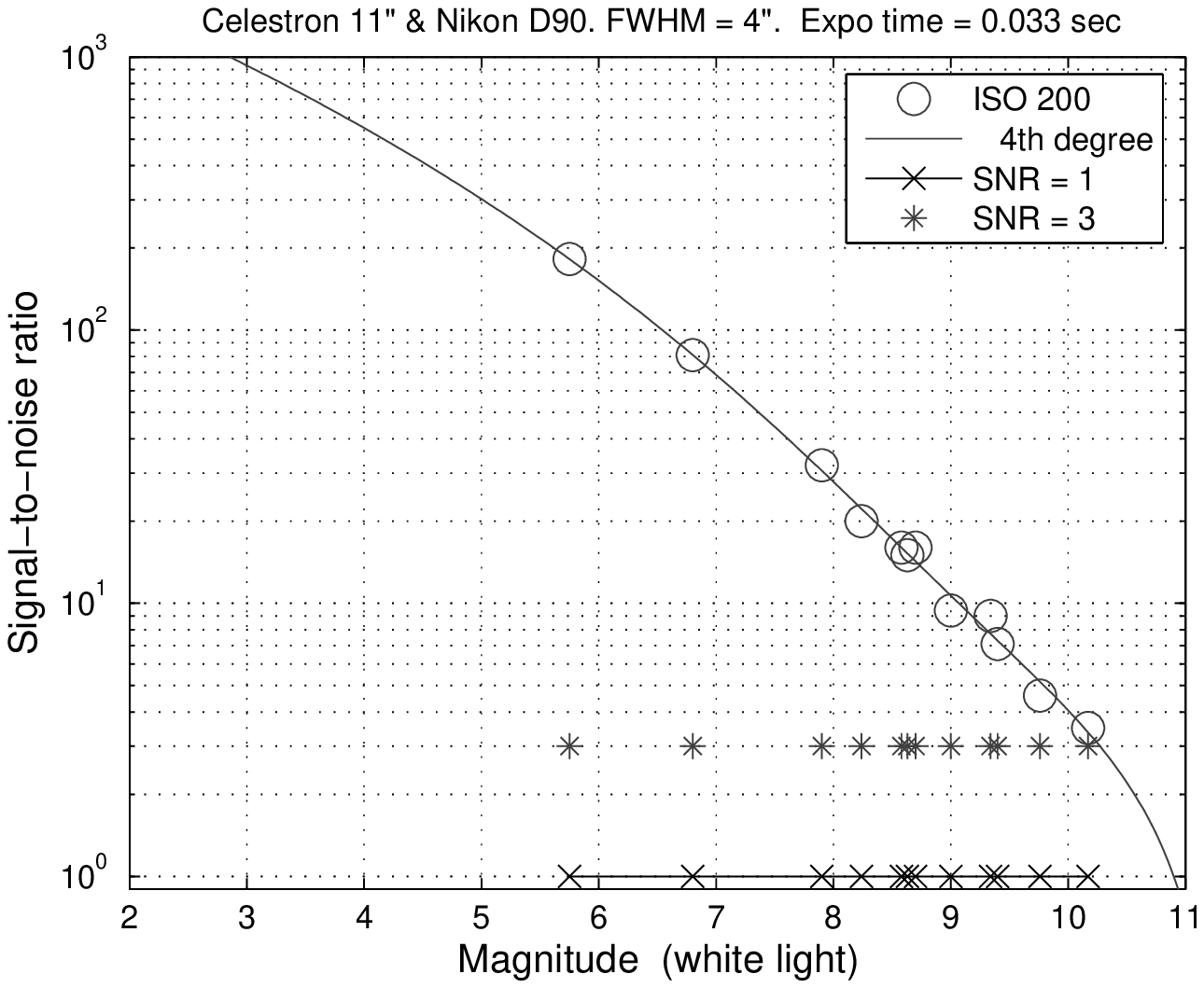,width = 1.0\linewidth} \caption{The
S/N ratio vs. B+G+R mag. The exposure time is 0.033 s.}\label{fig3}
\end{minipage}
\hfill
\begin{minipage}[t]{.45\linewidth}
\centering
\epsfig{file = 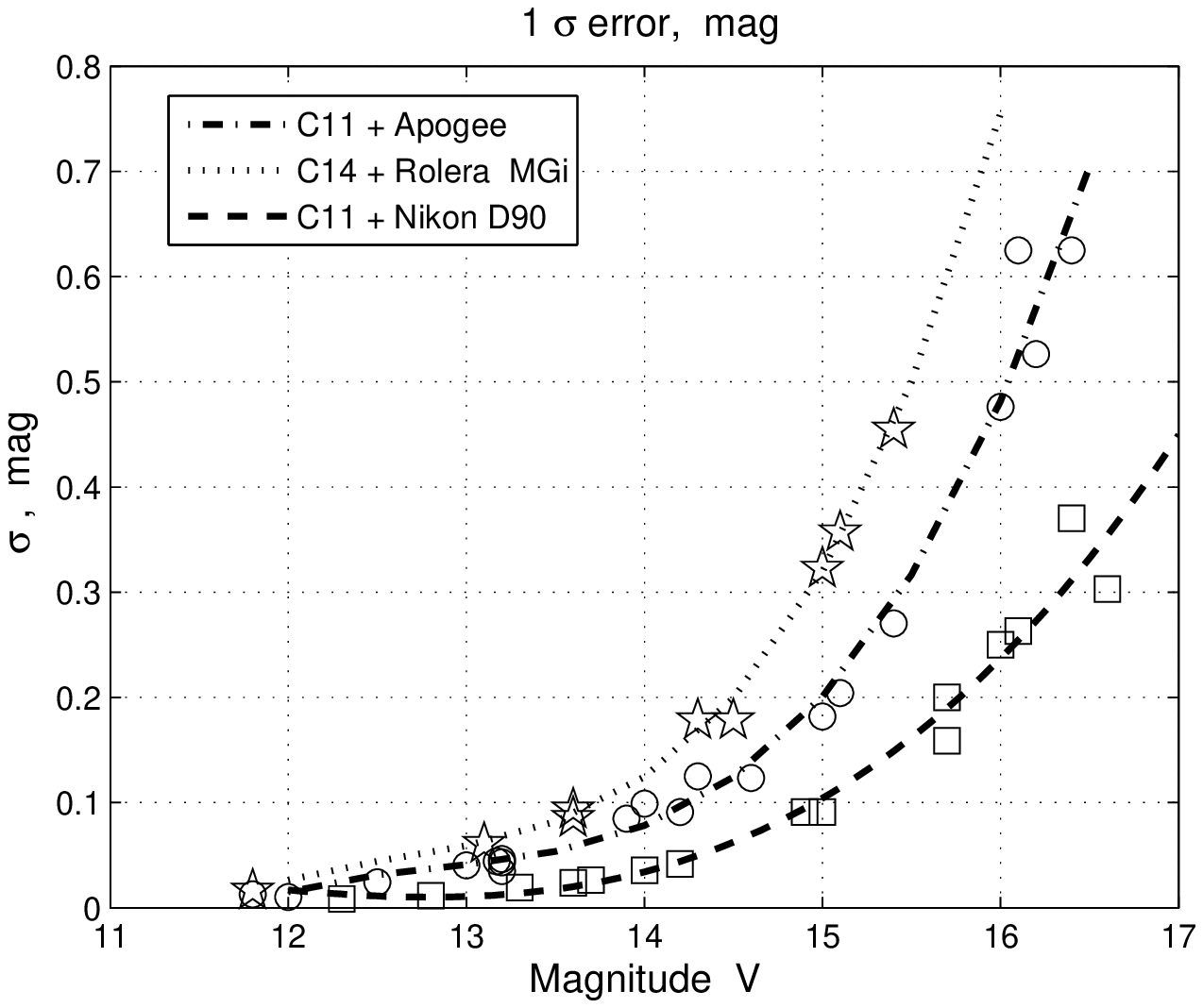,width = 1.0\linewidth} \caption{The
photometric errors vs. V mag for three cameras: APOGEE (circles),
Rolera MGi (stars), and Nikon D90 (squares).}\label{fig4}
\end{minipage}
\end{figure}

\begin{table}
 \centering
 \caption{The comparative data of the CCD cameras ALTA U42, FLI PL-1001E and CMOS Camera EOS 20D.}\label{tab1}
 \vspace*{1ex}
 \begin{tabular}{cccc}
  \hline
  Camera & Apogee ALTA U42 & FLI PL-1001E & Canon EOS 20D$^{1}$ \\
  \hline
  Image format, pixels & 2048 x 2048 & 1024 x 1024 & 3504 x 2336 \\
  Pixel size, microns &  13.5 & 24.0 & 6.4 \\
  Readout noise in electrons & 14.9 &  9 & 7.5 (ISO 400) \\
  Dark Current, e-/px/sec & 1 (- 20$^{\circ}$C) & 0.2 (- 45$^{\circ}$C) & 0.2 (+20$^{\circ}$C) \\
  Dynamic &    16; 12-bit &  16-bit & 12-bit \\
  Peak QE (550 nm) & $\sim $ 93\% & $\sim $ 70\% & $\sim $ 50\% \\
  Full Well in electrons & 100K & 500K &  $\sim $ 50K \\
  \hline
  & & & $^{1}$ \cite{clark12} \\
   \end{tabular}
\end{table}

\section*{\sc results and conclusions}

\indent \indent Fig. 1 shows a graph of the flux through pixel
depending on the magnitude V calculated for the Celestron 11"
telescope equipped with Nikon D90 camera, ISO = 200.  As can be seen
from the graph, we can measure the stars of V $\approx$ 18 in the
photon mode when the signal from the star becomes comparable to the
dark current $\sim$ 1electron/pixel/sec.

The processing of the observational data obtained at the observatory
Peak Terskol, confirmed results of our calculations. We can measure
the stars up to 14 mag with high accuracy, as shown in Fig. 2.

Fig. 3 shows the signal-to-noise ratio of Nikon D90 in serial
shooting mode depending on the white light magnitude for the
exposure time of 33 ms. One can see that the continuous shooting
allows us to study rapid variability of bright stars up to 7 mag
with high photometric precision and of 9 mag with acceptable
accuracy. This opens the way to study the rapid variability of the
large number of stars, among which there are many objects, which
gave the name to the prototypes of stellar variability.

Fig. 4 shows the comparative data obtained with the CCD cameras ALTA
U42, Rolera Mgi and CMOS camera Nikon D90 with the telescopes
Celestron 11" and 14". The exposure time is from 15 to 30 sec. The
one sigma error is proportional to the inverse value of the S/N
ratio. We can conclude from this graph that CMOS Nikon D90 provides
more accurate observations.

If we assume that the limiting stellar magnitude corresponds to the
signal-to-noise ratio equal to three, then these magnitudes in the V
band for the 11" telescope equipped with the cameras Rolera MGi,
Apogee ALTA E47 and Nikon D90 are 15.0, 15.5 and 16.5, respectively.
Note that the low dark sky background on Peak Terskol ($\sim$ 21.5
mag from square arc sec) enables long-term accumulation of the
signal. From this it is easy to estimate that the limiting magnitude
for the telescope Celestron 11" equipped with Nikon D90 camera will
be 19.5 magnitudes for one hour exposure time. Thus, the testing
results reveal that commercial CMOS cameras can create very serious
competition with modern CCD cameras in astronomy.

\section*{\sc acknowledgement}
\indent \indent
The authors would like to particularly thank Maxim Andreev at the
Peak Terskol Observatory for help during the observations.


\end{document}